\documentclass[runningheads]{llncs}

\usepackage{cite}
\usepackage{amsmath,amssymb,amsfonts}
\usepackage{graphicx}
\usepackage{textcomp}
\usepackage{xcolor}
\usepackage{booktabs}
\usepackage{mathtools}
\usepackage{latexsym,ifthen,url}
\usepackage{pgfplots}
\usepackage{tikz}
\usetikzlibrary{decorations.pathreplacing}

\usepackage{subfig}
\usepackage{xspace}
\usepackage{graphicx}
\usepackage[lambda,logic,notions,landau]{cryptocode}
\usepackage{multirow}
\usepackage{dashbox}
\usepackage{listings}
\usepackage{enumitem}
\usepackage{fancybox}
\usepackage[most]{tcolorbox}
\AtEndEnvironment{proof}{\phantom{}\qed}
\usepackage{makecell}
\usepackage{algorithm}
\usepackage{algpseudocode}
\usepackage[export]{adjustbox}
\usepackage{booktabs}
\usepackage{arydshln}
\usepackage[misc]{ifsym}
\usepackage{multirow}
\usepackage{siunitx}
\usepackage{amssymb}
\usepackage{orcidlink}

\newcommand{\remove}[1]{}

\newcommand{\dELDP}{\ensuremath{\mathsf{dELDP}}\xspace}
\newcommand{\etal}{\emph{et~al.}\xspace}

\newcommand{\vaeComp}{VAE\xspace}

\newcommand{\lape}{\ensuremath{\mathsf{PSY}}\xspace}

\newcommand{\PPL}{\textit{PPL}\xspace}

\newcommand{\extension}[1]{}

\begin{document}
\title{Short Paper: PSY: Posterior Sampling Based Privacy Enhancer in Large Language Models}
\titlerunning{Posterior Sampling Based Privacy Enhancer in LLM}
	
\author{Yulian Sun\inst{1,3,\orcidlink{0009-0007-0088-0385}} \and Li Duan\inst{1,2, \orcidlink{0009-0001-1663-2622}}
	 \and Yong Li\inst{1,\textrm{\Letter}}}

\institute{
	Huawei Technologies D\"{u}sseldorf, D\"{u}sseldorf, Germany\\ \email{\{yualian.sun1, li.duan, yong.li1\}@huawei.com}\\
	\and
	Paderborn University, Paderborn, Germany\\ \email{liduan@mail.upb.de}\\
	\and
	Ruhr University Bochum, Bochum, Germany \\ \email{ yulian.sun@edu.ruhr-uni-bochum.de}
	}
	
	\authorrunning{Y. Sun \and L. Duan \and Y. Li }
	
	\maketitle

        \begin{abstract}
            Privacy vulnerabilities in LLMs, such as leakage from memorization, have been constantly identified, and various mitigation proposals have been proposed. LoRA is usually used in fine-tuning LLMs and a good entry point to insert privacy-enhancing modules. In this ongoing research, we introduce \lape, a \textbf{P}osterior \textbf{S}ampling based Privac\textbf{Y} enhancer that can be used in LoRA.
            We propose a simple yet effective realization of \lape using posterior sampling, which effectively prevents privacy leakage from intermediate information and, in turn, preserves the privacy of data owners. We evaluate LoRA extended with \lape against state-of-the-art membership inference and data extraction attacks.
            The experiments are executed on three different LLM architectures fine-tuned on three datasets with LoRA.
            In contrast to the commonly used differential privacy method, we find that our proposed modification consistently reduces the attack success rate. Meanwhile, our method has almost no negative impact on model fine-tuning or final performance. Most importantly, \lape reveals a promising path towards privacy enhancement with latent space extensions.
        \end{abstract}
        
        \keywords{privacy enhancing technology, large language models}

        \section{Introduction}\label{sec1}
            Large Language Models (LLMs) mainly refer to transformer-based neural language models that contain up to hundreds of billions of parameters.
            Existing LLMs are firstly pre-trained on massive text data, such as LLaMA \cite{touvron2023llama} and GPT-4 \cite{achiam2023gpt}, then adapted for specific tasks.
            Up to 2020, direct fine-tuning was the mainstream method used for adaptation. However, to update a LLM with billions of parameters is computationally expensive. To deal with the problem, Hu \etal \cite{hu2021lora} propose LoRA, an efficient and computation-saving method in 2021.
            Instead of updating all LLM parameters during fine-tuning, LoRA freezes the base weights of the pre-trained model, then pays attention to a distinct updating matrix.
            
            Unfortunately, no matter which methods are used for fine-tuning, LLM may memorize information. At inference time, LLMs may even reproduce sensitive information about samples in the training and fine-tuning dataset when being attacked \cite{hintersdorf2024finding, nasr2023scalable, carlini2021extracting}. The threat becomes more severe when a LLM is released as a public service.
            
            \begin{figure}
                \centering
                \includegraphics[width=0.80\linewidth]{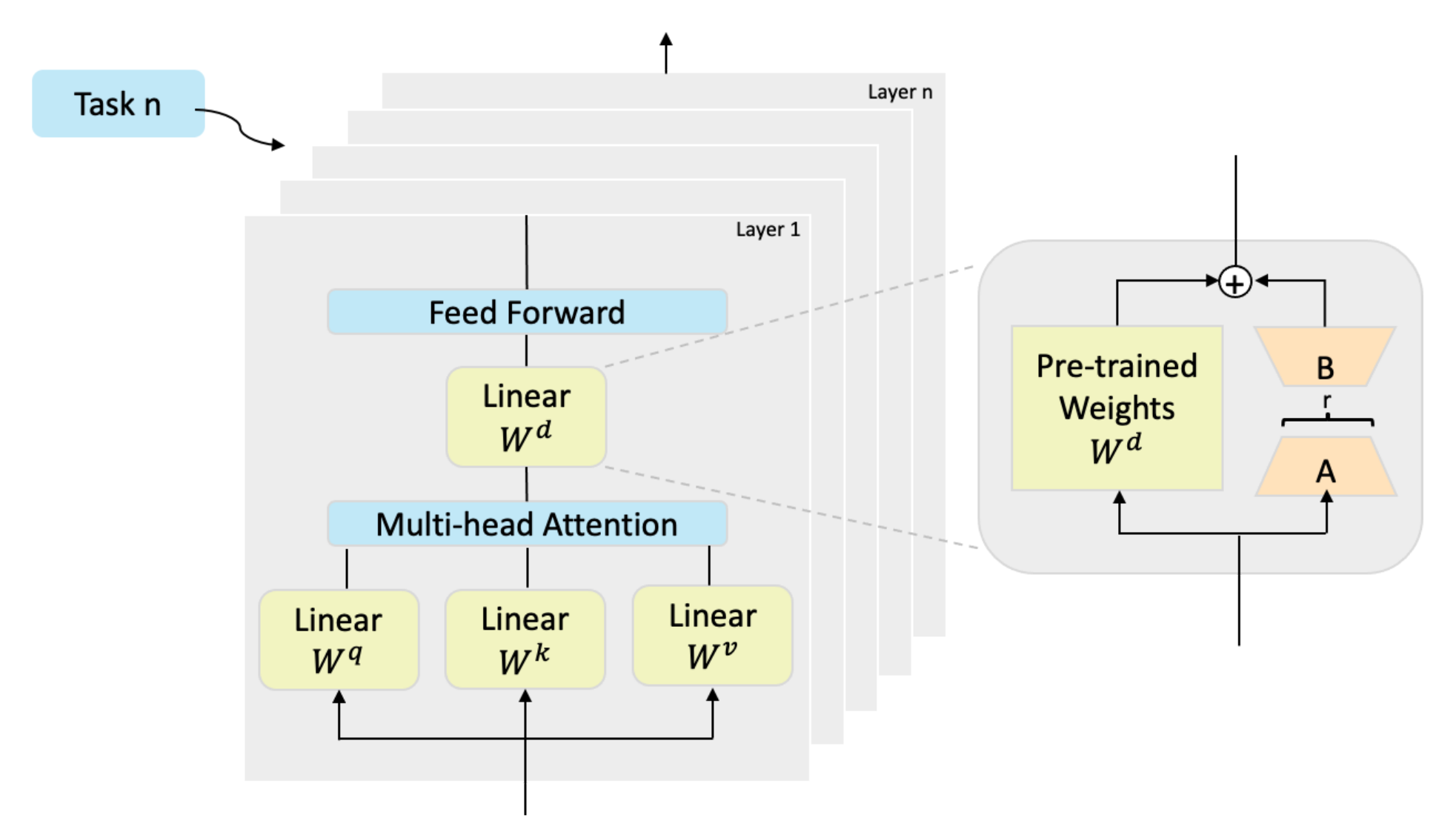}
                \caption{LoRA usage example, the location and structure of LoRA in a LLM for specific task. LoRA freezes the pre-trained model weights $W^d$ and simulates the amount of parameter changes through low-rank decomposition with $BA$, which greatly reduces the number of trainable parameters for downstream tasks.}
                \label{fig:lore-and-z}
            \end{figure}

        \subsection{Mitigating Leakage from Memorization}\label{sec:intro-mem-leakage}
            Various efforts have been made to mitigate this issue.
            The mitigation proposals include identifying and removing highly memorized samples \cite{ren2024unveiling}, 
            preventing the generation of sensitive samples \cite{somepalli2023understanding}, and the most recent solution proposed by Hintersdorf \etal in 2024: localizing and deactivating (removing) memorization neurons \cite{hintersdorf2024finding}. 
            Since modifying the structure of trained LLM as done Hintersdorf \etal \cite{hintersdorf2024finding} can defend against leakage caused by memorization, an interesting question to ask is:
            \textit{Instead of removing, can we \textbf{add} tiny modules to LLMs that can effectively hinder memorization?}

        \subsection{Hints from Defenses Against Related Attacks}
            By investigating other attacks related to memorization, we can find hints about the necessary properties of the modules we are looking for.
            
            Membership inference attacks (MIAs) aim to determine whether a given sample is in the training dataset \cite{shokri2017membership}.
            Besides being a privacy threat itself, vulnerability to MIA is a more general indicator of whether a trained model memorizes and leaks private information \cite{leino2020stolen, shokri2017membership, yeom2017unintended}.
            It was confirmed by Leino and Fredrikson \cite{leino2020stolen} in {\sf USENIX Security} 2020 that DP with a \textbf{small} $\epsilon \approx 0.25$, i.e., large perturbation, can thwart the membership inference attack caused by model memorization. However, the authors also pointed out the dilemma: large perturbation leads to sharp model accuracy reduction, whereas small perturbation ($\epsilon \approx 16$) makes little difference in front of the attacker.
            Therefore, it is preferred to have modules that can provide a DP-like guarantee, and its parameters can be adapted to the trained model and datasets.

            A review of structural defenses against gradient inversion \cite{huang2021evaluating} or data extraction attack (DEA) \cite{luo2021feature} attacks\footnote{Both attacks aim to recover sensitive information from different interfaces of the model.} can also help us locate the appropriate module.
            While being an efficiency enhancer in many machine learning models \cite{suh2016echo, xu2017variational}, variational auto-encoder (\vaeComp) has also been studied for its positive impact on privacy protection.
            In 2022, Scheliga \etal proposed PRECODE, a privacy-enhancing module based on \vaeComp which can be composed with model before training, and shown its effective defense against gradient inversion \cite{scheliga2022precode} in federated learning (FL).
            In 2023, Wang \etal \cite{wang2023differentially} injected controlled noise into recurrent VAE to generate synthesized data, which would then be used for FL training to preserve privacy.
            Furthermore, a recent study by Sun \etal \cite{sun2024exploiting} in 2024 has shown that posterior sampling layer trained in the model can have a variant of DP called distance-based empirical local DP ($\dELDP$) in inference.
            The experiments in \cite{sun2024exploiting} demonstrated that modules with \dELDP can mitigate feature reconstruction attacks effectively in FL.

        \subsection{Construbution}
        Thus, we propose a new module that can effectively address problems caused by memorization.
        More specifically, we make the following contribution in this short paper.
        \begin{itemize}
            \item We propose a tiny module called \textbf{P}osterior \textbf{S}ampling based Privac\textbf{Y} enhancer (\lape) which, when added into a LLM, can effectively alleviate the memorization by a LLM of the fine-tuning dataset.
            We also identify the optimal insertion point of \lape in LoRA (
            see Figure \ref{fig:vm}
            for an illustration).
            \item We empirically evaluate \lape on two state-of-the-art privacy attacks (MIA and DEA) on three Models trained on three real-world datasets and compare \lape with DP parameterized with a small $\epsilon$.
        \end{itemize}

        \section{Other Related Work}
        \paragraph{Membership Inference Attack (MIA).}
        MIAs are typical privacy attacks, which allow attackers to distinguish if a given data sample was used for training in a target model \cite{shokri2017membership}.  
        Meeus \etal \cite{meeus2023did} provide a document-level membership inference for real-world LLMs. Essentially, they seek to establish whether a LLM encountered a certain document during its training or not with the help of overfitting. 
        Fu \etal \cite{fu2023practical} proposed 
        a novel MIA approach based on self-calibrated probabilistic variation, which leverages memorization (which occurs before overfitting) rather than overfitting itself. This allows the adversary to collect a dataset with a similar distribution from public APIs. Both \cite{meeus2023did} and \cite{fu2023practical} assume that the adversary has access to samples closely resembling the original training data. To deal with limitations,
        Mattern \etal \cite{mattern2023membership} proposed a MIA using the concept of neighborhood comparison. This MIA compares model scores for a given sample to scores of synthetically generated neighbor texts and assumes that the attack has perfect knowledge about the training data distribution.

        \paragraph{Data Extraction Attack (DEA).}
        DEA refers to a method where an adversary gets individual training instances from a language model that has been developed on private datasets \cite{carlini2021extracting}.
        Carlini \etal \cite{carlini2021extracting} proposed a DEA from language models, which aims to recover individual training examples by querying the language model. The attack scenario is the first one that involves querying a LLM. Yu \etal \cite{yu2023bag} proposed benchmark tricks for improving training data extraction using a publicly available dataset. The authors explore both text generation (e.g., sampling strategy) and text ranking (e.g., token-level criteria) tricks.
        Nasr \etal \cite{nasr2023scalable} propose a DEA that focuses on extractable memorization. This method refers to training data that an adversary can efficiently extract by querying a LLM. Besides, the authors also demonstrate that the adversary can successfully extract gigabytes of training data from three various types of LLMs: open-source models, semi-open models and closed models.

        \paragraph{Defenses Using Structural and Internal Randomness.}
        In contrast to introducing external noise as in centralized DP or Local DP, Duan \cite{duan2009privacy} introduced the notion of privacy-without-noise in 2009, 
        which estimates the corresponding $(\epsilon, \delta)$ from the inherent randomness in the dataset and the query and highlights that non-uniform internal noise can also help achieve reasonable DP-like privacy.
        Duan's work also inspires \cite{burchard2019empirical, sun2024exploiting} that offers ways to empirically quantify the internal and structural randomness, and \cite{burchard2019empirical,sun2024exploiting} also proposed structures that are effective against privacy attacks.
        The idea of PRECODE \cite{scheliga2022precode} is to insert a full VAE, which also has internal randomness, into models such that after training, the models can have empirical defenses against gradient inversion.

        \section{Methodology}
        \subsection{\lape}
        	In this section, we introduce details of \lape, beginning with the general idea of adding a posterior sampling layer and following with the algorithmic description.  
        	\subsection{General Idea}
\begin{figure}[htbp]
            \centering
            \includegraphics[width=0.50\textwidth]{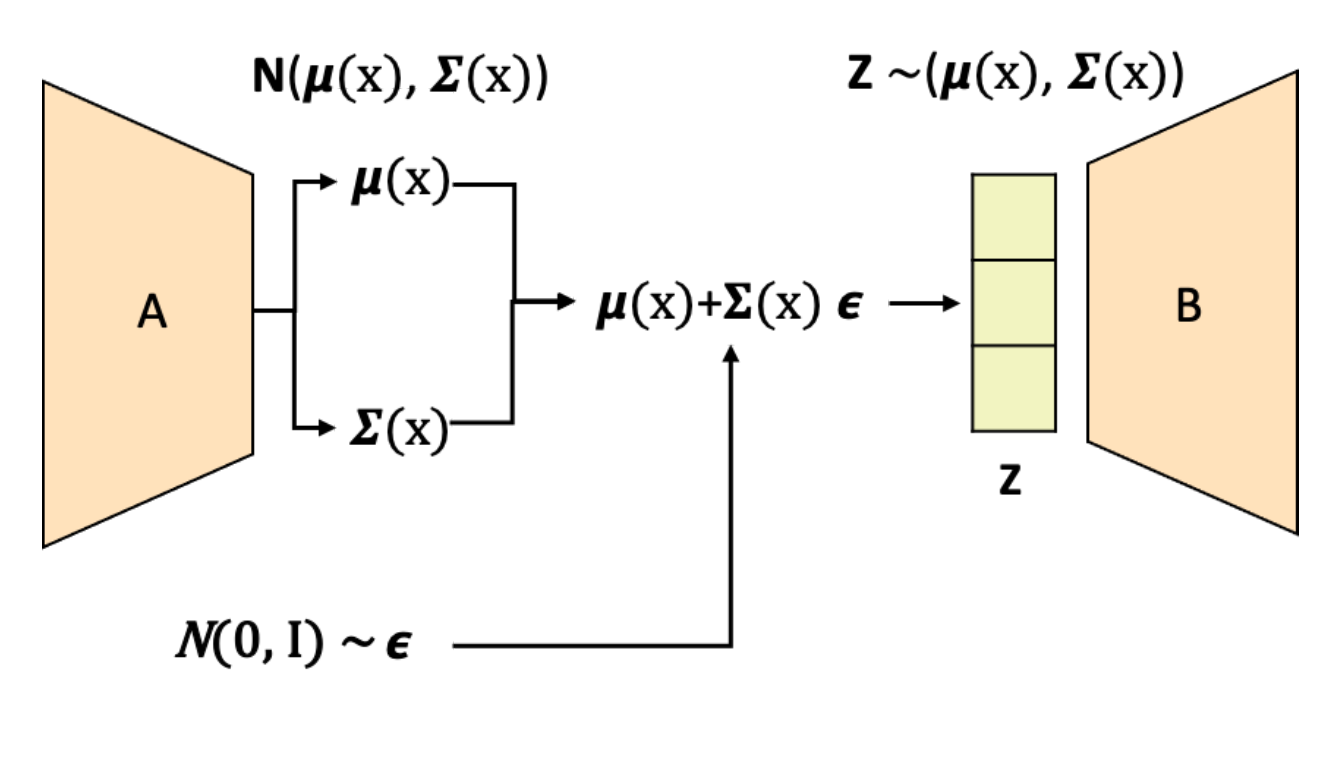}
            \caption{Structure of Posterior Sampling in between $A$ and $B$.}
            \label{fig:vm}
            \begin{flushleft}
            \end{flushleft}
        \end{figure}

        	We insert a posterior sampling layer in LoRA.
                \vaeComp contains two parts: encoder and decoder. The biggest difference between \lape and PRECODE is that \lape only needs the posterior sampling layer of VAE (in the encoder), i.e., the mapping from $x$ to the latent vector $z$. The posterior sampling layer is inserted in $A$ and $B$ (See Figure \ref{fig:vm}). Given the input $x$, the output of $A$ maps $x$ into a fixed term latent-vector $z$, which is sampled from $\mu$ and $\Sigma$. Thus, the latent-vector $z$ is
        
                \begin{equation}
                    z = \mu(Ax) + \epsilon \Sigma(Ax)
                \end{equation}
	
        Assuming a pre-trained model with weight $W_0\in \mathbb{R}^{d\times k}$, we constrain its updates by representing $\Delta W \in \mathbb{R}^{d\times r}$ with a low-rank decomposition $W_0 + \Delta W = W_0 + BA$, where $B \in \mathbb{R}^{d\times r}$, $A \in \mathbb{R}^{r\times k}$, and the rank $r \ll$ min$(d, k)$ \cite{hu2021lora}. $A$ and $B$ contain trainable parameters. During training, $W_0$ is frozen and does not receive gradient updates, while noting both $W_0$ and $\Delta W = BA$ are multiplied with the same input $x$, and their respective output vectors are summed coordinate-wise. For a hidden layer $h=W_0x$, then our modified forward pass yields:
        \begin{equation}
            h = W_0x + \Delta Wx = W_0x + B[\mu(Ax) + \epsilon \Sigma(Ax)] = W_0x + Bz
        \end{equation}

       Figure \ref{fig:vm} shows the details of integrating posterior sampling and LoRA. Here, as we referred to the original method of LoRA, we also use a Gaussian initialization for A and zero for B. Thus, $\Delta W= BA$ is viewed as zero from the start of training. In our method, we also keep the scaling of $\Delta Wx$ by $\frac{\alpha}{r}$, where $\alpha$ is a constant in $r$. \lape also keeps efficiency, since tuning $\alpha$ is equivalent to tuning learning rate \cite{hu2021lora}.

        \section{Experiments}
        \subsection{Setup and Baselines}
         We evaluate our method on three open-source LLMs, including \textbf{GPT-Neo} (2.7B) \cite{black2021gpt}, \textbf{RedPajama-INCITE} (2.8B) \cite{pedpajama-v1}
         and \textbf{Pythia} (6.9B) \cite{biderman2023pythia}, which have from 2.7 to 6.9 Billion of parameters. These LLMs are pre-trained and we download them from Hugging Face \footnote{https://huggingface.co/EleutherAI/gpt-neo-2.7B}\footnote{https://huggingface.co/togethercomputer/RedPajama-INCITE-Chat-3B-v1}\footnote{https://huggingface.co/EleutherAI/pythia-6.9B}. We then fine-tune these models on other three datasets, including \textbf{AG's News} \cite{Zhang2015CharacterlevelCN}, \textbf{Twitter} \cite{go2009twitter} and \textbf{Wikitext-103} \cite{merity2016pointer} (See Table \ref{tab:dataset_label}).
        
        \begin{table}
            \centering
            \caption{Datasets and Models. B represents Billion.}
            \label{tab:dataset_label}
            \scalebox{0.77}{
            \begin{tabular}{@{}lclcc@{}}
                \toprule
                \textbf{Model} & \textbf{$\#$Param.} & \textbf{$\#$Dataset} &
                \textbf{$\#$Samples} &
                \textbf{$\#$Labels} \\
                \midrule
                \textbf{GPT-Neo} & 2.7B & \textbf{AG's News} & 120K & 4\\
                \textbf{RedPajama-INCITE} & 2.8B & \textbf{Twitter} & 300k & 2\\
                \textbf{Pythia} & 6.9B & \textbf{WikiText-103} & 200k & -\\
                \bottomrule
            \end{tabular}
            }
        \end{table}
    
  We evaluate our methods from both functionality and defensive ability perspectives. From a functionality perspective, we compare our method with a baseline that does not use protection methods or classic differential privacy techniques.
  We fine-tune our method and baseline with \textbf{GPT-Neo} and \textbf{RedPajama-INCITE} for 4 epochs. We set batch size as 8 and leaning rate as $1\mathrm{e}{-4}$ on 2 GPUs. For \textbf{Pythia}, we fine-tune for 2 epochs. Besides, we set batch size as 4 and keep the learning rate as $1\mathrm{e}{-4}$ on 4 GPUs.

  As a defense comparing, we choose differential privacy (DP) from open-source library \textit{pyvacy} \cite{pyvacy} with a small $\epsilon=0.1$, i.e. large perturbation. The fine-tuning settings are the same as the baselines.

  From a defensive ability perspective, we evaluate the above fine-tuned models on two state-of-the-art privacy attacks: a data extraction attack (DEA) and a membership inference attack (MIA).

  DEA needs no access to the trained model under the black-box setting. The implemented DEA refers to \cite{carlini2021extracting}. The adversary can efficiently extract by querying the trained model without any prior knowledge of the training data. We choose models which are fine-tuned on \textbf{AG's News}.
  Then, we evaluate DEA with these fine-tuned models.

  Similar to DEA, MIA needs no access to the trained model, either. We implement MIA under the \textit{black-box} setting, which provides a strong adversary and an upper bound of privacy loss. The implemented MIA refers to \cite{mattern2023membership}, with a proposal model to generate data. The adversary knows nothing about the trained model but only accesses by querying, We evaluate MIA on three datasets.

\subsection{Results}


We analyze the results with perplexity (\PPL), which measures the fine-tuned models in its (word) predictions. We score each sample based on the ratio of the perplexity after
and before lowercasing the original text, we name this ratio as \textit{Lowercasing}. We use \textit{Lowercasing} as a measurement because it drastically alters the perplexity of memorized content, which requires a specific lowercase. Besides, we also score the ratio of zlib \cite{zlib} entropy and the perplexity, we call this ratio as \textit{Zlibbing}. Zlib, as a text compressor, can detect repeated patterns, so \textit{Zlibbing} dramatically alters the perplexity of memorized content, which has repeated texts. We choose the top-1 result of each metric, our method has similar perplexity compared with the baseline and DP methods. However, our method achieve lower ratio both in \textit{Lowercasing} and \textit{Zlibbing}, which means our method generates efficient content and not potentially memorize training samples.
There is almost no difference between the different models when comparing baseline, DP, and our method with \PPL. The results are all close to 1.1. For \textit{Lowercasing}, our method almost has lower values compared with DP. Specifically, in \textbf{RedPajama-INCITE} and \textbf{Pythia} on \textbf{AG's News}, our method achieves 3.062 and 7.458. However, in \textbf{GPT-Neo} on \textbf{Wikitext-103}, our method achieves the highest value of 42.127. For \textit{Zlibbing}, our method almost achieves lower values than DP (See Table \ref{tab:DEA}).

We refer to \cite{mattern2023membership} to analyze the results of MIA. We describe attack performances according to their true positive rates (TPR) under very low false positive rates (FPR) by different threshold values. We choose 10\%, 1\% and 0.1\% as our target FPR values. The results on different models show that our method has a lower TPR compared with the baseline, which means our method is effective to defend MIA. 
In comparison with DP, our method almost gets lower values. However, in \textbf{RedPajama-INCITE} and \textbf{Pythia} on \textbf{AG's News}, DP has better results than our method. The values are even zero when the ratios are $1\%$ and $0.1\%$ (See Table \ref{tab:Memebership Inference Attack}).
 
\begin{table*}
    \centering
    \caption{\lape against DEA}
    \label{tab:DEA}

\scalebox{0.77}{
\begin{tabular}{*{13}{c}}
\toprule
  \multicolumn{3}{c}{\multirowcell{2}{\textbf{Score}}}
  &  & \multicolumn{3}{c}{\textbf{AG's News}}
  & \multicolumn{3}{c}{\textbf{Twitter}}
  & \multicolumn{3}{c}{\textbf{Wikitext-103}} \\
\cmidrule(r){4-13}

  \multicolumn{1}{c}{}
  & \multicolumn{1}{c}{}
  & \multicolumn{1}{c}{}
  &  & \PPL
  & \textit{Lowercasing}
  & \textit{Zlibbing}
  & \PPL
  & \textit{Lowercasing}
  & \textit{Zlibbing}
  & \PPL
  & \textit{Lowercasing}
  & \textit{Zlibbing}
  \\
\midrule
  \multicolumn{2}{c}{\textbf{Baseline}}
  & \multirowcell{3}{\textbf{GPTNeo}}
  &  & 1.030 & 37.387 & 18388.217 & 1.039 & 36.066 & 15226.839 & 1.016 &  23.143 & 16881.580\\

\multicolumn{2}{c}{\textbf{DP}}
 &  &  & 1.016 & 39.331 & 14243.051 & 1.016 & 37.708 & 14165.225 & 1.016 &  36.46 & 13244.788 \\

  \multicolumn{2}{c}{\textbf{Ours}}
  &  &  & 1.016 & 37.762 & 14165.508 & 1.016 & 37.239 & 14094.263 & 1.015 &  42.127 & 13942.004 \\
\midrule
  \multicolumn{2}{c}{\textbf{Baseline}}
  & \multirowcell{3}{\textbf{RedPajama}}
  &  &  1.136 & 11.347 & 3906.395 & 1.546 & 4.353 & 994.626 & 1.069 & 1.575 & 522.252 \\

  \multicolumn{2}{c}{\textbf{DP}} 
  &  &  & 1.133 & 11.788 & 3115.580 & 1.135 & 11.345 & 3914.205 & 1.112 & 12.280 & 3086.298 \\

  \multicolumn{2}{c}{\textbf{Ours}}
  &  &  & 1.182 & 3.062 & 1543.762 & 1.217 & 8.116 & 2742.597 & 1.456 & 3.348 & 1494.253 \\
\midrule
  \multicolumn{2}{c}{\textbf{Baseline}}
  & \multirowcell{3}{\textbf{Pythia}}
  &  & 1.049 & 15.701 & 11835.779 & 1.065 & 20.959 & 10082.962 & 1.100 & 4.038 & 2577.540 \\

  \multicolumn{2}{c}{\textbf{DP}}
  &  &  & 1.052 & 15.260 &  11106.952 & 1.049 & 15.406 & 11848.253 & 1.068 & 19.128 & 9424.102 \\

  \multicolumn{2}{c}{\textbf{Ours}}
  &  &  & 1.077 & 7.457 & 8368.949 & 1.056 & 20.768 & 10983.864 & 1.100 & 9.836 & 6732.477 \\
\bottomrule
\end{tabular}
}
\end{table*}

        \begin{table*}
            \centering
            \caption{\lape against MIA under different FPR Metrics}
            \label{tab:Memebership Inference Attack}
            \scalebox{0.89}{
            \begin{tabular}{*{13}{c}}
            \toprule
          \multicolumn{3}{c}{\multirowcell{2}{\textbf{False Positive Ratio}}}
          &  & \multicolumn{3}{c}{\textbf{AG's News}}
          & \multicolumn{3}{c}{\textbf{Twitter}}
          & \multicolumn{3}{c}{\textbf{Wikitext-103}} \\
            \cmidrule(r){4-13}
            
              \multicolumn{1}{c}{}
              & \multicolumn{1}{c}{}
              & \multicolumn{1}{c}{}
              &  & \textbf{10\%}
              & \textbf{1\%}
              & \textbf{0.1\%}
              & \textbf{10\%}
              & \textbf{1\%}
              & \textbf{0.1\%}
              & \textbf{10\%}
              & \textbf{1\%}
              & \textbf{0.1\%}
              \\
            \midrule
          \multicolumn{2}{c}{\textbf{Baseline}}
          & \multirowcell{3}{\textbf{GPTNeo}}&
          & 36.23\% & 14.57\% & 12.83\% & 42.37\% & 4.30\% & 3.20\%  & 31.53\% & 1.3\% & 0.86\% \\

        \multicolumn{2}{c}{\textbf{DP}}
        & & & 36.50\% & 1.43\% & 1.03\% & 11.13\% & 1.67\% & 0.33\% & 39.50\% & 1.43\% & 1.03\% \\

            \multicolumn{2}{c}{\textbf{Ours}}
          &  &  & 25.73\% & 0.33\% & 0.10\% & 11.16\% & 0.26\% & 0.16\%  & 37.66\% & 9.63\% & 6.36\% \\
            \midrule
          \multicolumn{2}{c}{\textbf{Baseline}}
          & \multirowcell{3}{\textbf{RedPajama}}
          &  &  36.03\% & 11.03\% & 9.80\% & 21.33\%  & 0.93\% & 0.7\% & 41.50\% & 18.23\% & 16.10\% \\

         \multicolumn{2}{c}{\textbf{DP}} 
         &  &  & 8.1\% & 0.0 & 0.0 & 7.10\% & 0.10\% & 0.07\% & 41.90\% & 8.33\% & 7.60\%  \\

          \multicolumn{2}{c}{\textbf{Ours}}
          &  &  & 26.20\% & 8.73\% & 7.93\% & 7.26\% & 0.1\% &0.06\% & 37.60\% & 9.83\% & 6.86\% \\
        \midrule
          \multicolumn{2}{c}{\textbf{Baseline}}
          & \multirowcell{3}{\textbf{Pythia}}
          &  & 50.90\% & 10.57\% & 10.03\% & 94.80\% & 47.50\% & 39.27\% & 52.50\% & 12.36\% & 11.83\% \\
        
         \multicolumn{2}{c}{\textbf{DP}}
         &  &  & 15.67\% & 0.0 & 0.0 & 7.20\% & 0.13\% & 0.10\% & 43.10\% & 3.76\% & 3.36\% \\
        
          \multicolumn{2}{c}{\textbf{Ours}}
          &  &  & 26.21\% & 3.40\% & 2.66\% & 10.70\% & 0.50\% & 0.26\% & 37.66\% & 9.63\% & 6.36\%\\
        \bottomrule
        \end{tabular}
        }
        \end{table*}
        
        \section{Conclusion and Future Work}\label{sec12}
        In this short paper, we propose an efficient module \lape to mitigate memorization-related attacks, and the empirical results are mainly positive.
        On the one hand, the current encourages us to investigate whether other LoRA-based fine-tuning methods, e.g., AdaLoRA \cite{zhang2023adalora} and QLoRA \cite{dettmers2024qlora} can be composed with \lape and ensures privacy protection.
        On the other hand, we would like to quantify the privacy guarantee of \lape further and explore alternative modules that can have similar effects.

        \section{Acknowledgements}
        This research is fully funded by the European Research Center of Huawei Technologies. We thank Ricardo Mendes for the valuable discussion. We thank anonymous reviewers for the various constructive comments and suggestions.
	
	\bibliographystyle{splncs04}
	\bibliography{sn-bibliography}
	
\end{document}